\documentclass[preprint,notoc]{JHEP3}
\usepackage{epsfig}

\def\eslt{\not\!\!{E_T}}
\def\to{\rightarrow}

\def\bi{\begin{itemize}}
\def\ei{\end{itemize}}

\def\tst{\tilde t}
\def\ttau{\tilde \tau}

\def\tg{\tilde g}

\def\tell{\tilde\ell}

\def\tw{\widetilde W}
\def\tz{\widetilde Z}
\def\alt{\stackrel{<}{\sim}}
\def\agt{\stackrel{>}{\sim}}
\def\be{\begin{equation}}  
\def\ee{\end{equation}}  

\title{Indirect, Direct and Collider Detection of\\
Neutralino Dark Matter in the mSUGRA Model}

\author{Howard Baer, Alexander Belyaev, 
Tadas Krupovnickas and Jorge O'Farrill 
\\ Department of Physics, Florida State University Tallahassee, FL 32306, USA\\
E-mail: \email{baer@hep.fsu.edu},\email{belyaev@hep.fsu.edu}
\email{tadas@hep.fsu.edu},\email{ofarrill@hep.fsu.edu}}

\preprint{\vbox{\hbox{FSU-HEP-040520}}} 

\abstract{
We examine the prospects for supersymmetry discovery in the minimal 
supergravity (mSUGRA) model via indirect detection of neutralino
dark matter. We investigate rates for muon detection in
neutrino telescopes, and detection of photons, positrons and
anti-protons by balloon and space based experiments. 
We compare the discovery 
reach in these channels with the reach for direct detection of dark matter,
and also with the reach of collider experiments such as Fermilab Tevatron, 
CERN LHC and a $\sqrt{s}=0.5-1$ TeV linear $e^+e^-$ collider.
We pay particular attention to regions of model parameter space in accord with
recent WMAP results on the dark matter density of the universe. 
We find that 3rd generation
direct dark matter detection experiments should be able to cover
the {\it entire} WMAP allowed portion of the 
hyperbolic branch/focus point (HB/FP) region of parameter space, while
the IceCube neutrino telescope can cover almost all this region. This is
in contrast to the case of the CERN LHC or a linear $e^+e^-$ collider, 
where only a fraction of the HB/FP region can be accessed.
In addition, we show that detection of $\gamma$s, $e^+$s and $\bar{p}$s 
should occur
in much of the HB/FP region, as well as in 
the low $m_{1/2}$ portion of the $A$ annihilation funnel, and will be
complementary to searches via colliders in these regions.
}


\begin{document}

\section{Introduction}
\label{sec:intro}

Evidence for cold dark matter (CDM) in the universe comes from 
observations of galactic rotation curves and binding of galaxies in clusters,
from matching observations of large scale structure with simulations, 
from gravitational microlensing, from the baryonic density of the universe as 
determined by Big Bang nucleosynthesis, from observations of supernovae in 
distant galaxies, and from measurements of anisotropies in the cosmic microwave
background radiation (CMB)\cite{review_cdm}. 
In particular, the 
Wilkinson Microwave Anisotropy Probe~(WMAP)\cite{wmap} collaboration 
has extracted a variety of cosmological parameters from fits to
precision measurements of the CMB radiation.
The properties of a flat universe in the 
standard $\Lambda CDM$ cosmological model 
are characterized by 
the density of baryons ($\Omega_b$), matter density 
($\Omega_m$), vacuum energy 
($\Omega_\Lambda$) and the expansion rate ($h$) which are measured to be:
\begin{eqnarray}
\Omega_b&=&0.044\pm 0.004 \\
\Omega_m&=&0.27\pm 0.04   \\
\Omega_\Lambda&=&0.73\pm 0.04 \\
h&=&0.71^{+0.04}_{-0.03} .
\end{eqnarray}
From the WMAP results, a value for the cold dark matter 
density of the universe can be derived:
\begin{equation}
\Omega_{CDM}h^2=0.1126^{+0.0081}_{-0.0090}
( {^{+0.0161}_{-0.0181}} )  \mbox{ at 68(95)\% CL}.
\end{equation}
While the origin of dark energy in the universe remains a conundrum, 
there exists a number of hypothetical 
candidate elementary particles to fill the role of CDM.

A particularly attractive candidate for CDM is the lightest neutralino in
$R$-parity conserving supersymmetric models\cite{goldberg}.
In the paradigm minimal supergravity (mSUGRA) model\cite{msugra}, 
it is assumed that
at the scale $Q=M_{GUT}$, there is a common scalar mass $m_0$, 
a common gaugino mass $m_{1/2}$, and a common trilinear term $A_0$.
The soft SUSY breaking terms can be calculated at scale $Q=M_{weak}$
via renormalization group evolution.
Electroweak symmetry breaking occurs radiatively (REWSB) due to the
large top quark mass, so that the
bilinear soft breaking term $B$ can be traded for the weak scale ratio
of Higgs vevs $\tan\beta$, and the magnitude (but not the sign) of
the superpotential $\mu$ term can be specified. Thus, the mSUGRA
model is characterized by four parameters plus a sign choice:
\be
m_0,\ m_{1/2},\ A_0,\ \tan\beta ,\ {\rm and}\ sign(\mu ) .
\ee
Once these model parameters are specified, then all sparticle
masses and mixings are determined, and scattering cross sections
may be reliably calculated.

In the early universe at very high temperatures, 
the lightest neutralino $\tz_1$ will be in thermal equilibrium, so that 
its number density is well determined. As the universe expands and cools,
the expansion rate outstrips the neutralino interaction rate, and 
a relic density of neutralinos is frozen out.
The neutralino relic density $\Omega_{\tz_1}h^2$ at the present time 
can be determined by solving the Boltzmann equation for
neutralinos in a Friedmann-Robertson-Walker universe. 

In most of the parameter space of the mSUGRA model, 
it turns out that a value of
$\Omega_{\tz_1}h^2$ well beyond the WMAP bound is generated. Only certain 
regions of the mSUGRA model parameter space give rise to a relatively low
value of $\Omega_{\tz_1}h^2$ in accord with cosmological measurements 
and theory. These regions consist of:
\begin{enumerate}
\item The bulk annihilation region at low values of $m_0$ and $m_{1/2}$,
where neutralino pair annihilation occurs at a large rate via $t$-channel
slepton exchange.
\item The stau co-annihilation region at low $m_0$ where 
$m_{\tz_1}\simeq m_{\ttau_1}$ so that $\tz_1$s may co-annihilate
with $\ttau_1$s in the early universe\cite{stau_co}.
\item The hyperbolic branch/focus point (HB/FP) region\cite{hb_fp} 
at large $m_0$
near the boundary of the REWSB excluded region where $|\mu |$ becomes
small, and the neutralinos have a significant higgsino component, 
which facilitates annihilations to $WW$ and $ZZ$ pairs\cite{fmw,bb2}.
\item The $A$-annihilation funnel, which occurs at very large 
$\tan\beta\sim 45-60$\cite{Afunnel}. 
In this case, the value of $m_A\sim 2m_{\tz_1}$.
An exact equality of the mass relation isn't necessary, since
the $A$ width can be quite large ($\Gamma_A\sim 10-50$ GeV);
then $2m_{\tz_1}$ can be several widths away from resonance,
and still achieve a large $\tz_1\tz_1\to A\to f\bar{f}$ annihilation
cross section. The heavy scalar Higgs $H$ also contributes to 
the annihilation cross section.  
\end{enumerate}
In addition, there exists a region of neutralino top-squark
co-annihilation\cite{stop} 
(for very particular $A_0$ values) and a light Higgs
$h$ annihilation funnel\cite{bb} (at low $m_{1/2}$ values).

In past years, the bulk annihilation region of parameter space 
was favored. This situation has changed in that the 
low $m_0$ and $m_{1/2}$ portion of the bulk annihilation
region has been excluded by LEP2 chargino and Higgs search bounds, 
while the larger $m_0$ and $m_{1/2}$ portion generally predicts 
values of $\Omega_{\tz_1}h^2$ beyond the rather restrictive
upper bound of $\Omega_{\tz_1}h^2 <0.129$ obtained from WMAP.
Any remaining portions of the bulk region
give rise to large- usually anomalous- predictions of
the rate for $BF(b\to s\gamma )$ decays and muon anomalous magnetic moment
$a_\mu =(g-2)_\mu/2$\cite{bbb,sug_chi2}. 
An increase of either of the parameters $m_0$
or $m_{1/2}$ leads generally to heavier sparticle masses and $m_h$
values, so that predictions for loop induced processes become
more SM-like.

A panoply of collider and non-accelerator experiments are now operating
or will soon be deployed that will shed light on CDM.
Prospects for detecting dark matter and determining its properties
are particularly bright in the case of neutralinos from supersymmetry.
Neutralino dark matter may well be produced at 
the Fermilab Tevatron $p\bar{p}$ collider\cite{tev}, the CERN LHC\cite{lhc} 
$pp$ collider, and a $\sqrt{s}=0.5-1$ TeV linear $e^+e^-$ 
collider\cite{nlc}. In addition, there exist
both direct and indirect non-accelerator 
dark matter search experiments that are 
ongoing and proposed. Direct dark matter detection has been recently 
examined by many authors\cite{direct}, and observable signal rates
are generally found in either the bulk annihilation region, or in the
HB/FP region, while direct detection of DM seems unlikely in the 
$A$-funnel or in the stau co-annihilation region.

Indirect detection of neutralino dark matter\cite{eigen} may occur via
\begin{enumerate}
\item observation of high energy neutrinos originating from
$\tz_1\tz_1$ annihilations in the core of the sun or earth\cite{neut_tel}, 
\item observation of $\gamma$-rays originating from neutralino annihilation
in the galactic core or halo\cite{gamma} and 
\item observation of positrons\cite{positron} or anti-protons\cite{pbar}
originating from neutralino annihilation in the galactic halo. 
\end{enumerate}
The latter
signals would typically be non-directional due to the influence of galactic
magnetic fields, unless the neutralino annihilations occur relatively
close to earth in regions of clumpy dark matter.

The indirect signals for SUSY dark matter have been investigated 
in a large number of papers, and computer codes which yield
the various signal rates are available\cite{neutdriver,darksusy}. 
Recent works find that the various indirect signals occur at 
large rates in the now disfavored bulk annihilation region, 
and also in the HB/FP region\cite{fmw}.
Naively, this is not surprising since the same regions of parameter space
that include large neutralino annihilation cross sections in the early
universe should give large annihilation cross sections as sources of
indirect signals for SUSY dark matter. 
In Ref. \cite{bo}, it was pointed out that the $A$ annihilation
funnel can give rise to large rates for cosmic $\gamma$s, $e^+$s and
$\bar{p}$s. However, neutralino-nucleon scattering cross sections 
are low in the $A$ annihilation funnel, so that 
no signal is expected at neutrino telescopes, which depend more on the
neutralino-nucleus scattering cross section than on the neutralino
annihilation rates.
Our goal in this paper is to combine the projected discovery contours 
from Tevatron, LHC and LC searches with those of direct and indirect 
dark matter search experiments. It turns out that each distinct region
of mSUGRA parameter space which gives rise to an acceptable
$\Omega_{\tz_1}h^2$ value also gives a set of unique predictions for
combinations of collider and non-accelerator experiments.

We begin our analysis by generating 
sparticle mass spectra using Isajet v7.69\cite{isajet}, which includes 
full one-loop radiative corrections
to all sparticle masses and Yukawa couplings, 
and minimizes the scalar potential using the
renormalization group improved 1-loop effective potential 
including all tadpole contributions, evaluated at an 
optimized scale choice which accounts for leading two loop terms.
Good agreement between $m_h$ values is found in comparison with
the FeynHiggs program, and there is good agreement as well in the 
$m_A$ calculation between Isajet and SoftSUSY, Spheno and Suspect codes,
as detailed in Ref. \cite{kraml}. To evaluate the indirect signals expected
from the mSUGRA model, we adopt the DarkSUSY 3.14 
package\cite{darksusy} interfaced
to Isasugra\footnote{Isasugra is a subprogram of the Isajet package that 
calculates sparticle mass spectra and branching fractions for a variety of
supersymmetric models}.
For our calculation of the neutralino relic density, we use the
IsaReD program\cite{bbb_rd} interfaced with Isajet. 
IsaReD calculates all relevant neutralino pair annihilation and co-annihilation
processes with relativistic thermal averaging\cite{gg}.
An important element of the calculation is that IsaReD 
calculates the neutralino
relic density using the Isajet $t$, $b$ and $\tau$ Yukawa couplings
evaluated at the scale $Q=m_A$. 
The Yukawa coupling 
calculation begins with the $\overline{DR}'$ fermion masses at
scale $Q=M_Z$, and evolves via 1-loop 
SM renormalization group equations (RGEs) to
the scale $Q_{SUSY}=\sqrt{m_{\tst_L} m_{\tst_R}}$, 
where complete MSSM 1-loop threshold
corrections are implemented. Evolution at higher mass scales is
implemented via 2-loop MSSM RGEs. The final RGE solution is gained after
iterative running of couplings and soft terms 
between $M_Z$ and $M_{GUT}$ and back until
a convergent solution is achieved.

The remainder of this paper is organized as follows. In 
Sec.~\ref{exp_overview}, we present an overview of indirect, direct and 
collider search experiments for neutralino dark matter. In Sec. \ref{results},
we discuss the impact of different halo models on our calculations. We then
present our main results, which are a series of plots in the
$m_0\ vs.\ m_{1/2}$ plane of the mSUGRA model showing the comparative
reach of various indirect, direct and collider searches for neutralino
dark matter. In Sec. \ref{conclude}, we present our conclusions.

\section{Experimental overview}
\label{exp_overview}

\subsection{Neutrino telescopes}

A novel technique for detecting dark matter is to search for neutrino
signals coming from neutralino annihilation in the core of the earth or the
sun\cite{neut_tel}. 
As the sun or earth proceeds on its orbital path, neutralinos can 
be swept up and become gravitationally captured by the process of 
neutralino-nucleon scattering until the recoil neutralino
velocity drops below the escape velocity. 
Thus, a high density of neutralinos can accumulate in the core of the earth or
sun, where they can efficiently annihilate. Neutralino 
annihilation into SM final
states such as $b\bar{b}$, $c\bar{c}$, $W^+W^-$ or $ZZ$ ultimately yields 
neutrinos via the $b\to c\ell\nu_\ell$, $c\to s\ell\nu_\ell$, 
$W\to\ell\nu$ or $Z\to\nu_\ell\bar{\nu}_\ell$ decays. 
The neutrinos can propagate out of the core of the earth or sun, 
and be detected via $\nu_\mu\to \mu$ conversions in 
neutrino telescopes such as Antares or IceCube. 
In fact, limits have already been obtained by Amanda for the case of
neutralino annihilation in the core of the earth\cite{amanda}.

The Antares $\nu$ telescope
should be sensitive to $E_\mu >10$ GeV; it is in the process of deployment 
and is expected to turn on in 2006\cite{antares}.
It should attain a sensitivity of $100-1000$ $\mu$s/km$^2$/yr. 
The IceCube $\nu$ telescope is also in the
process of deployment at the south pole\cite{icecube}. 
It should be sensitive to
$E_\mu >25-50$ GeV, and is expected to attain a sensitivity of 
$40-50$ $\mu$s/km$^2$/yr. Full deployment of all detector elements is
expected to be completed by 2010.

The rate for
neutralino annihilation in the sun or earth is given by
\be
\Gamma_A ={1\over 2}C\tanh^2(t_{\odot}/\tau),
\ee
where $C$ is the capture rate, $A$ is the total annihilation rate times
relative velocity per volume, $t_\odot$ is the present age of the solar
system and $\tau =1/\sqrt{CA}$ is the equilibration time. 
For the sun, the age of the solar system exceeds the 
equilibration time, so $\Gamma_A\sim \frac{C}{2}$, and the muon
flux tends to follow the neutralino-nucleon scattering rate rather than the
neutralino pair annihilation cross section. 
In this case, the indirect dark matter detection rate should be 
relatively independent of the dark matter halo profile, aside from 
the value of the local neutralino relic density. 
Thus, uncertainties in the
predicted rates for neutrino detection via neutralino annihilation in the 
core of the sun should be low.
In contrast to the sun, the earth has typically a much
longer equilibration time $\tau$, so that $\Gamma_A\sim\frac{1}{2}C^2 A t^2$,
and is hence more sensitive to the neutralino annihilation cross section
times relative velocity. Muon rates from neutralino annihilation in the 
core of the earth are typically much lower than those from the sun. In 
addition, expected rates from the earth may be diminished even further
by solar depletion effects: see J. Lundberg and J. Edsj\"o, 
Ref. \cite{lundberg} (these effects are not included in DarkSUSY 3.14,
but are included in DarkSUSY 4.0).

\subsection{Detection of $\gamma$s} 

Neutralinos may also collect in the core of the galaxy, where they can 
annihilate at a high rate. In this case, 
$\tz_1\tz_1\to q\bar{q},\ W^+W^-,\ ZZ\to hadrons$ which gives 
rise to photons typically from $\pi^0\to\gamma\gamma$ decay. 
It is also possible for $\tz_1\tz_1\to \gamma\gamma$ (or $Z\gamma$), 
in which case $E_\gamma\simeq m_{\tz_1}$. The signature is spectacular 
in this case, but the rates are loop suppressed. 
The $\gamma$ rays 
can be detected down to sub-GeV energies with space-based
detectors such as EGRET\cite{egret} or GLAST\cite{glast}. 
Ground based arrays require much higher
photon energy thresholds of order $20-100$ GeV. Experiments such as 
GLAST should be sensitive to rates of order $10^{-10}$ $\gamma$s/cm$^2$/sec
assuming $E_\gamma >1$ GeV. 
In fact, it has recently been suggested that the extra-galactic gamma ray 
background radiation as measured by EGRET is well fit by a model
of neutralino annihilation\cite{elsasser}; see also \cite{deboer}.
It is important to note that the prediction 
for rates for $\gamma$ detection depends sensitively on models for the 
neutralino density near the galactic core. The latter quantity is poorly known,
so there can be a wide range in predicted rates depending on 
assumptions about the galactic halo profile.

\subsection{Detection of $e^+$s} 

Cosmic positrons may also be searched for from neutralino annihilations in the
galactic halo. In this case, the positrons would arise as decay products of
heavy quarks, leptons and gauge bosons produced in neutralino annihilations.
Space based anti-matter detectors such as Pamela\cite{pamela} 
and AMS-02\cite{ams} will be able 
to search for anomalous positron production from dark matter annihilation.
The cosmic positron excess as measured by HEAT\cite{heat} 
has been suggested as 
having a source in galactic halo neutralino annihilations\cite{baltz,deboer}.
It is suggested by Feng {\it et al.}\cite{fmw} 
that a reasonable observability criteria
is that 
signal-to-background ($S/B$) rates 
should be greater than the $1-2\%$ level.
To calculate the
$S/B$ rates, we adopt fit C from Ref. \cite{fmw} for the  
$E^2 d\Phi_{e^+}/d\Omega dE$ background rate:
\be
E^2 d\Phi_{e^+}/d\Omega dE =1.6\times 10^{-3}\ E^{-1.23},
\ee
where $E$ is in GeV.
We compute the signal
using the DarkSUSY positron flux evaluated at an ``optimized'' energy
of $E=m_{\tz_1}/2$, as suggested in Ref. \cite{fmw}. A $S/B\sim 0.01$ 
rate may be detectable\cite{fmw,eigen} 
by experiments such as Pamela and AMS-02.

\subsection{Detection of $\bar{p}$s} 

Antiprotons may also be produced in the debris of neutralino annihilations 
in the galactic halo. Such antiprotons have been measured by the 
BESS collaboration\cite{bess}.
The differential flux of antiprotons from
the galactic halo, $d\Phi_{\bar{p}}/dE_{\bar{p}}d\Omega$, 
as measured by BESS, has a peak in the kinetic energy distribution 
at $E_{\bar{p}}\sim 1.76$ GeV.
The height of the peak
at $E_{\bar{p}}\sim 1.76$ GeV is $\sim 2\times 10^{-6}$ 
${\bar{p}}$/GeV/cm$^2$/s/sr. Signal rates in the range
of $10^{-7}-10^{-6}$ ${\bar{p}}$/GeV/cm$^2$/s/sr might thus
provide a benchmark for observability.  

\subsection{Direct search for neutralino DM}

If indeed all space is filled with relic neutralinos, 
then it may be possible to 
directly detect them via their scattering from nuclei.
Early
limits on the spin-independent neutralino-nucleon cross-section 
($\sigma_{SI}$) have 
been obtained by the CDMS\cite{cdms}, EDELWEISS\cite{edelweiss} 
and ZEPLIN1\cite{zeplin} groups, while a signal was 
claimed by the DAMA collaboration\cite{dama}. 
Collectively, we will refer to the reach from these groups as the ``Stage 1''
dark matter search. Depending on the neutralino mass,
the combined limit on $\sigma_{SI}$
varies from  $10^{-5}$ to $10^{-6}$~pb. This cross section
range is beyond the predicted levels from most supersymmetric models.
However, experiments in the near future
like CDMS2, CRESST2\cite{cresst}, ZEPLIN2 and EDELWEISS2 
(Stage 2 detectors) should have a 
reach of the order of 
$10^{-8}$~pb.
In fact, the first results from CDMS2 have recently appeared, and yield
a considerable improvement over the above mentioned Stage 1 
results\cite{cdms2}.
Finally, a number of experiments such as GENIUS\cite{genius}, 
ZEPLIN4\cite{zeplin4} and XENON\cite{xenon} are in 
the planning stage. We refer to these as Stage 3 detectors, which promise 
impressive limits of the order of $\sigma_{SI}<10^{-9}$ -- $10^{-10}$~pb, and 
would allow the exploration of a considerable part of parameter space of many 
supersymmetric models. In particular, the Stage 3 direct DM detectors
should be able to probe almost the entire HB/FP region of mSUGRA model 
parameter space. We note here in addition 
that the Warm Argon Program (WARP)\cite{rubbia}
promotes a goal of detecting neutralino-nucleus scattering cross sections 
as low $10^{-11}$ pb.

\subsection{Fermilab Tevatron}

The Fermilab Tevatron $p\bar{p}$ collider can search for neutralino dark 
matter by detecting superparticle production reactions which lead 
to anomalous missing energy in the final state. For mSUGRA model
parameter choices in accord with bounds on the chargino mass 
from LEP2 ($m_{\tw_1}>103.5$ GeV), the most promising discovery channel
is the clean trilepton plus $\eslt$ final state\cite{old3l}, which typically
originates from $p\bar{p}\to\tw_1\tz_2 X\to 3\ell +\eslt +X$, where
$X$ stands for assorted hadronic debris, and where the trileptons
tend to originate from $\tw_1\to\ell\nu_\ell\tz_1$ and 
$\tz_2\to\ell\bar{\ell}\tz_1$ decay. Trilepton signal rates and backgrounds
have been calculated in Ref. \cite{trilep} using cuts SC2. These
results were updated and extended to large $m_0$ regions of parameter space  
in Ref. \cite{bkt}.

\subsection{CERN LHC}

The CERN LHC can search for neutralino dark matter by detecting 
superparticle production reactions which lead 
to anomalous missing energy in the final state. For LHC, however, 
gluino and squark production reactions are expected to be
the dominant SUSY cross sections. 
The gluinos and squarks can decay
through possibly lengthy cascade decays so that signal events will
consist of multi-jets plus isolated leptons plus $\eslt$\cite{cascade}. 
Signal and 
background levels have been computed in Ref. \cite{lhcsusy} for various
combinations of jet and lepton cuts. We adopt the most recent calculations of
Baer {\it et al.}\cite{lhcsusy} for our projections of the LHC reach, 
assuming 100 fb$^{-1}$ of integrated luminosity.

\subsection{Linear $e^+e^-$ collider}

A linear $e^+e^-$ collider operating at CM energy $\sqrt{s}=0.5-1$ TeV 
can also search for neutralino dark matter by detecting 
superparticle production reactions which lead 
to anomalous missing energy in the final state. The ultimate reach limits 
in the mSUGRA model depend on which sparticles are kinematically 
accessible to production. The reach contours are determined by\cite{bmt,bbkt}
\begin{itemize}
\item $e^+e^-\to \tell^+\tell^-$ at low $m_0$, followed typically by
$\tell\to\ell\tz_1$ decay and
\item $e^+e^-\to\tw_1^+\tw_1^-$ production at moderate to high $m_0$
values, where typically $\tw_1\to f\bar{f}'\tz_1$ or $W\tz_1$ ($f$ is any
$SM$ fermion). 
\item In intermediate regions, $e^+e^-\to \tz_1\tz_2$ and/or
$e^+e^-\to ZH,\ Ah$ may be accessible.
\end{itemize}

In the HB/FP region, the superpotential $\mu$ 
parameter becomes small, and the $\tw_1$ and $\tz_1$ become
increasingly higgsino like, and nearly mass degenerate.
In the small mass gap region, conventional cuts oriented towards a 
substantial $m_{\tw_1}-m_{\tz_1}$
mass gap must be replaced by new cuts. Ultimately, an $e^+e^-$ LC should
be able to see chargino pairs essentially up to the kinematic limit 
for their production, over all of mSUGRA parameter space. 
Reach plots have been presented in Ref. \cite{bbkt} 
assuming 100 fb$^{-1}$ of integrated luminosity for both the
$\sqrt{s}=0.5$ TeV and 1 TeV machines.

\section{Direct, indirect and collider searches in the mSUGRA model}
\label{results}

In this section, we evaluate the reach of various indirect, direct
and collider searches for neutralino dark matter in the mSUGRA
model.
For our predictions of indirect neutralino detection rates, we use the 
DarkSUSY program\cite{darksusy}, modified to interface with
Isajet v7.69. The Isajet subprogram Isasugra is used to predict the 
sparticle mass spectrum and decay widths for various supersymmetric
models, including mSUGRA.

\subsection{Calculational overview and dependence on halo model}

While predictions for the reach of colliders for neutralino dark matter
are firmly grounded in perturbative quantum field theory, many of the
predictions for direct and indirect detection of dark matter 
depend on the assumed density profile of the galactic halo.

For halo model dependence in the distribution of dark matter, we adopt
the default DarkSUSY value: 
a spherically symmetric isothermal distribution given 
by
\be
\rho (r)= \rho_0\frac{(r/r_0)^{-\gamma}}{(1+(r/a)^\alpha)^{\frac{\beta -\gamma}
{\alpha}}}(1+(r_0/a)^\alpha)^{\frac{\beta-\gamma}{\alpha}}
\ee
where $(\alpha ,\beta ,\gamma )=(2,\ 2,\ 0)$, $r_0=8.5$ kpc is the distance 
of earth to the galactic center, $\rho_0 =0.3$ GeV/cm$^{3}$ is the local 
dark matter density and $a=3.5$ kpc is a distance scale.
A number of other halo profiles are available in DarkSUSY, 
including
\begin{itemize}
\item Navarro, Frenk, White profile with $a=20$ kpc, 
with $(\alpha,\ \beta,\ \gamma )=(1,\ 3,\ 1)$\cite{frenk},
\item Moore {\it et al.} profile with $a=28$ kpc, 
with $(\alpha,\ \beta,\ \gamma )=(1.5,\ 3,\ 1.5)$\cite{moore} and
\item Kravtsov {\it et al.} profile with $\rho_0=0.6$ GeV/cm$^3$, $a=10$ kpc, 
with $(\alpha,\ \beta,\ \gamma )=(2,\ 3,\ 0.4)$\cite{kravtsov}.
\end{itemize}
The alternative distributions all use $r_0=8.0$ kpc as the sun
galactocentric distance value.
\FIGURE[htb]{
\epsfig{file=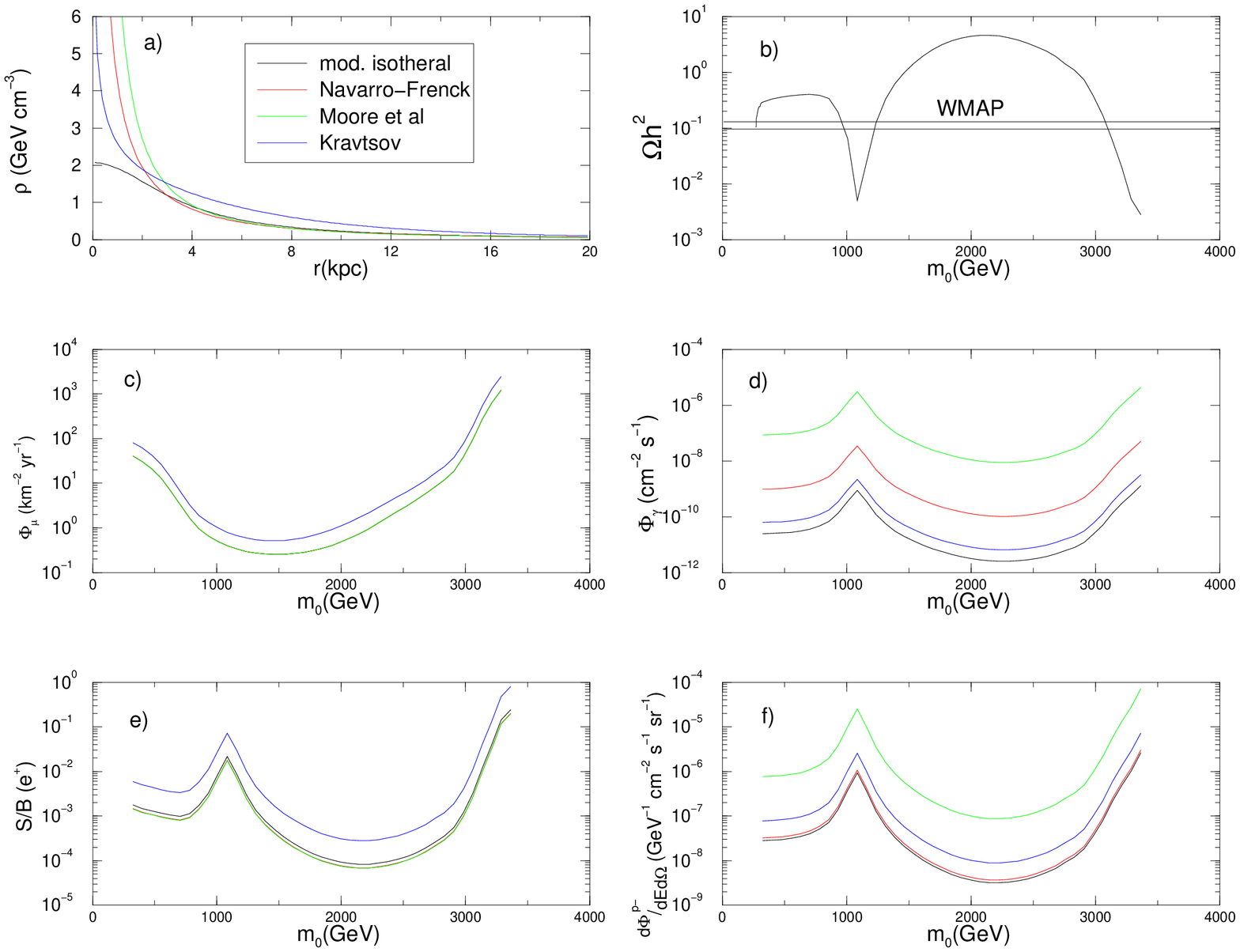,width=11cm,angle=-90} 
\caption{\label{halo}
In frame {\it a}), we show various halo model density profiles versus
galactic radial distance $r$ (kpc), while in
frame {\it b}), we show the neutralino relic density, along with the WMAP
bound (horizontal line). 
Frame {\it c}) shows the flux of muons from the sun.
In frames {\it d}), {\it e}), and {\it f}) we show rates for detection 
of $\gamma$s, $e^+$s and $\bar{p}$s from neutralino annihilations in the 
galactic core and halo, for different halo model choices. 
Frames {\it b})--{\it f}) are plotted versus $m_0$ for
$m_{1/2}=550$ GeV, $A_0=0$, $\tan\beta =50$ and $\mu <0$.}}

In Fig. \ref{halo}, we show in frame {\it a}) the various halo 
model distributions.
While the distributions are in close accord at $r=8.5$ kpc 
(the distance of the sun to the galactic center), 
the profiles disagree strongly as $r\to 0$, 
reflecting our relative ignorance of the density of dark matter
expected near the galactic center. 
In frame {\it b}), we show the neutralino relic density 
versus mSUGRA parameter $m_0$, with $m_{1/2}=550$ GeV, $A_0=0$, $\mu <0$ and
$\tan\beta =50$. The horizontal line shows the WMAP upper bound on
$\Omega_{\tz_1}h^2$. In frame {\it c}), we show the flux of muons
with $E_\mu >25$ GeV originating from neutralino annihilation to
neutrinos in the core of the sun. These curves only depend on the
{\it local} neutralino relic density, and not on the global 
galactic halo profile. Thus, the first three halo profiles yield the same
predictions for the muon detection rate, while the Kravtsov distribution, 
with a higher value for the local relic density, also gives higher 
rates for neutrino telescopes.
In frame {\it d}), we show the flux of
photons $\Phi_\gamma$ ($\gamma$s/cm$^2$/sec) 
with $E_\gamma >1$ GeV emanating from 
the galactic center, within a cone of solid angle 0.001 sr. 
The rates are large at large $m_0$ in the HB/FP region\cite{fmw}, 
and are also large
at $m_0\sim 1000$ GeV, where neutralino annihilation in the galactic core
can occur through the $A$ and $H$ resonances\cite{bo}. Overall, the 
predicted rates at fixed $m_0$ vary over several orders of magnitude, 
reflecting the different model predictions for the DM density at 
the galactic center. In frame {\it e}) , we show as well
the expected positron signal-to-background ($S/B$) rate for positrons
originating in neutralino annihilations in the galactic halo.
We see that observable rates may again occur in the HB/FP
region, and also in the $A$-annihilation funnel.
However, the uncertainty in the predictions due to variation in the halo model
is less severe than in the $\gamma$ case, since now the positrons 
are expected to arise relatively nearby in the galaxy, where the DM density
is much more constrained.
In frame {\it f}), we show the 
differential flux of antiprotons from
the galactic halo, $d\Phi_{\bar{p}}/dE_{\bar{p}}d\Omega$, 
for $E_{\bar{p}}=1.76$ GeV, which is near the location of the peak 
in the $\bar{p}$ kinetic energy distribution as measured 
by the BESS collaboration\cite{bess}. 
We see from the figure that
the largest rates 
occur in the HB/FP region, and also in the $A$-annihilation funnel; 
in these regions, the signal rates 
can extend into the region of observability.
Again, the predicted rates are sensitive to the assumed halo model.
We note, however, that the DarkSUSY default halo distribution tends to give
the most conservative of rate predictions, and it is possible that
indirect detection rates could be much higher than those shown using the 
default halo model. 

One final qualifying note: it is common practice to
rescale direct or indirect detection rates by a factor
$\Omega_{\tz_1}h^2/(\Omega_{CDM} h^2)_{ref}$ when 
$\Omega_{\tz_1}h^2<(\Omega_{CDM} h^2)_{ref}$, where $(\Omega_{CDM} h^2)_{ref}$
is some reference value $\sim 0.025-0.1$ which would give rise to the assumed
local dark matter density $\rho =0.3$ GeV/cm$^3$. We do not apply this practice
here. It is possible in a variety of non-standard cosmological models
(such as those containing quintessential scalar fields or 
primordial anisotropies) to obtain a dark matter relic density much higher 
than that obtained in a $\Lambda CDM$ model. See {\it e.g.} 
Ref. \cite{profumo} for details.

\subsection{Results for the mSUGRA model}

Our first results for the mSUGRA model are shown in Fig. \ref{fig10p} in the 
$m_0\ vs.\ m_{1/2}$ plane for $\tan\beta =10$, $A_0=0$ and $\mu >0$.
We take $m_t=175$ GeV. The left-most red region is excluded because
the stau becomes the LSP (in violation of search limits for 
stable charged or colored relics from the Big Bang), while the right-most 
red region is excluded due to a lack of REWSB.
The lower yellow region is excluded by LEP2 searches for chargino 
pair production ($m_{\tw_1}>103.5$ GeV), while the region below the 
yellow contour is excluded by LEP2 Higgs searches ($m_h>114.4$ GeV for a 
SM-like Higgs boson). The green shaded region has 
$\Omega_{\tz_1}h^2<0.129$, in accord with the upper bound on CDM from
WMAP. The left-most green strip along the low $m_0$ excluded region is the 
stau co-annihilation corridor, while the right-most green region
corresponds to the HB/FP region. A remaining green region sitting just 
atop the LEP2 excluded region is the light Higgs annihilation corridor, 
where $2m_{\tz_1}\sim m_h$\cite{bb}. 
The uncolored regions all give too large a
CDM relic density, and are thus excluded. 

The Fermilab Tevatron reach
contour corresponding to a $5\sigma$ signal for 10 fb$^{-1}$ of integrated
luminosity is denoted by TEV, while the CERN LHC $5\sigma$ reach 
for 100 fb$^{-1}$ is denoted LHC. The reach of a 
$\sqrt{s}=500$ GeV and 1000 GeV linear collider is denoted by LC500 and LC1000,
respectively, assuming 100 fb$^{-1}$ for each. 
We see from the figure that the Tevatron can cover the light Higgs 
annihilation corridor, while the LC1000 and LHC can cover the stau
co-annihilation region. The LHC can cover the HB/FP region up to
$m_{1/2}\sim 700$ GeV, which corresponds to a reach in $m_{\tg}$ of about
1.8 TeV. In this region, squarks and sleptons are
4-7 TeV in mass, and effectively decoupled from LHC production.
LHC can generate an observable signal cross section provided
$m_{\tg}$ is light enough. If $m_{\tg}$ is too heavy ($\agt 1.8$ TeV), 
then $pp\to\tg\tg X$ occurs at too low of a rate. Charginos and neutralinos
can be quite light in the HB/FP region since $|\mu |$ is small, but
their signal events are difficult for LHC to extract from background, owing
in part to a decreasing $m_{\tw_1}-m_{\tz_1}$ mass gap as $|\mu |$
decreases. The LC500 and LC1000 can still generate $\tw_1^+\tw_1^-$ pairs
in the HB/FP region if they are kinematically accessible, and should be able to
extract the corresponding low energy release events above 
SM background\cite{bbkt}. In the HB/FP region, we have the unusual situation
that the LC reach exceeds that of the CERN LHC. However, the high
$m_{1/2}$ portion of the HB/FP region gives rise to cases where
the chargino mass is too heavy to be accessible by a 
$\sqrt{s}=1$ TeV LC, so a thorough search for SUSY by colliders in this 
DM allowed region apparently can't be made 
(unless a higher energy LC is constructed).

We also show contours of 
\begin{itemize}
\item Stage 3 direct detection experiments
($\sigma_{SI}>10^{-9}$ pb; black contour),
\item reach of IceCube $\nu$ telescope with 
$\Phi^{sun}(\mu )=40\ \mu$s/km$^3$/yr and $E_\mu >25$ GeV
(magenta contour),
\item the $\Phi (\gamma)=10^{-10}\ \gamma$s/cm$^2$/s contour
with $E_\gamma >1$ GeV in a cone of 0.001 sr directed at the
galactic center (dark blue contour),
\item the $S/B>0.01$ contour for halo produced positrons
(blue-green contour) and
\item the antiproton flux rate 
$\Phi (\bar{p})=3\times 10^{-7}\ \bar{p}$s/cm$^2$/s/sr 
(lavender contour).
\end{itemize}
As noted by Feng {\it et al.}\cite{fmw}, 
{\it all} these indirect signals
are visible inside some portion of the HB/FP region, while {\it none} 
are visible in generic DM disallowed regions (under the assumed smooth
halo profiles).
The intriguing point is that almost the entire HB/FP region 
(up to $m_{1/2}\sim 1400$ GeV) can be explored
by the cubic km scale IceCube $\nu$ telescope! It can also be explored 
(apparently at later times) by the Stage 3 direct DM detectors. 
Given the relative time scales of the various search
experiments, if SUSY lies within the upper HB/FP region, then
it may well be discovered first by IceCube (and possibly Antares),
with a signal being later confirmed by direct DM detection and possibly the 
TeV scale linear $e^+e^-$ collider. There is also some chance to obtain
indirect $\gamma$, $e^+$ and $\bar{p}$ signals in this region.
Notice that if instead SUSY lies within the stau co-annihilation corridor, 
then it will be easily discovered by the LHC (for $\tan\beta =10$), but all
indirect detection experiments will find null results in their DM searches.

\FIGURE[htb]{
\epsfig{file=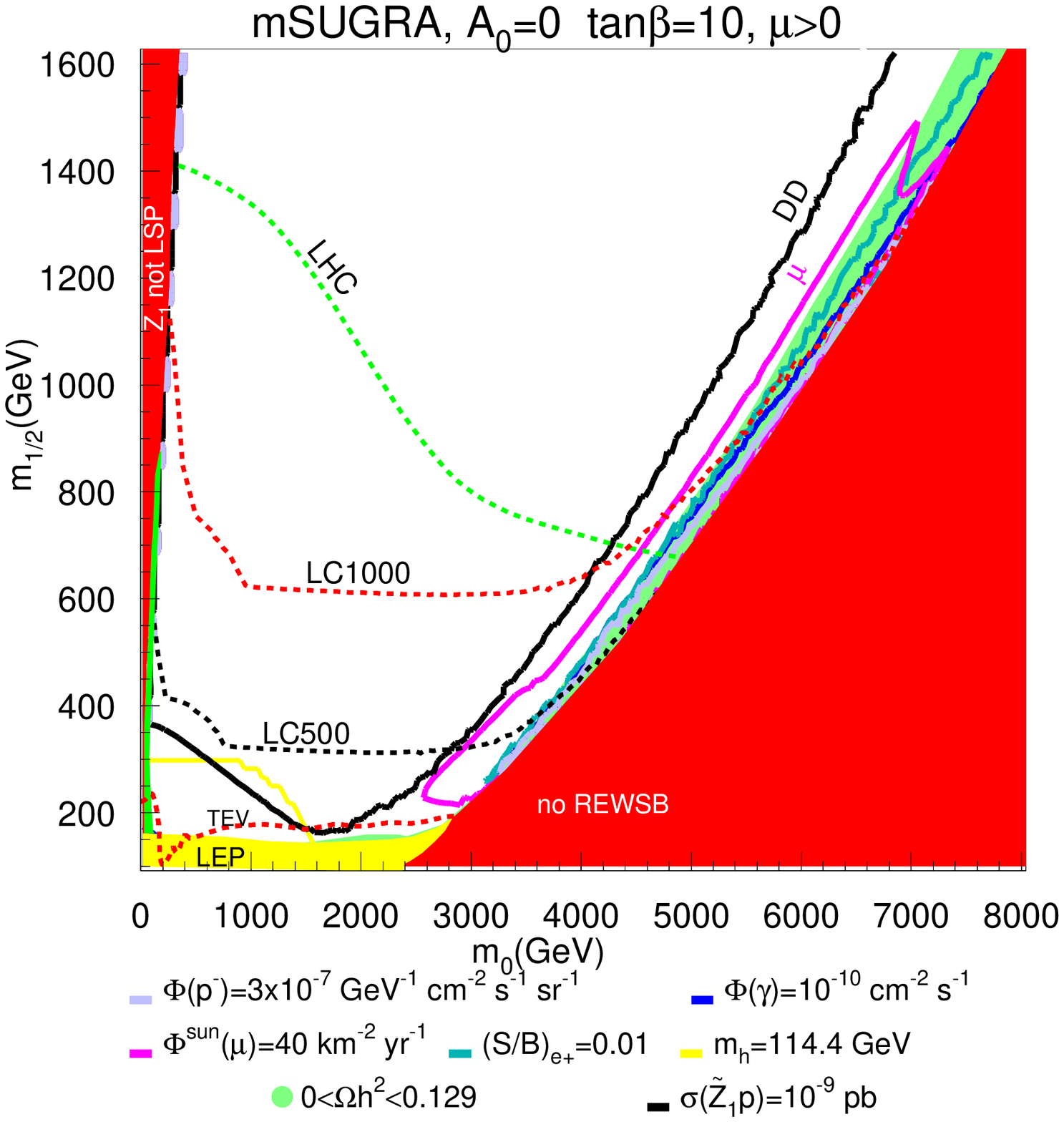,width=16cm} 
\caption{\label{fig10p}
A plot of the reach of direct, indirect and collider searches for 
neutralino dark matter in the $m_0\ vs.\ m_{1/2}$ plane, for 
$A_0=0$, $\tan\beta =10$ and $\mu >0$.}}

In Fig. \ref{fig30p}, we show again the $m_0\ vs.\ m_{1/2}$ plane, 
but this time for $\tan\beta =30$ (other parameters remain the same).
The results are rather similar to the $\tan\beta =10$ case shown in
Fig. \ref{fig10p}, although the stau annihilation corridor has expanded 
somewhat to higher $m_{1/2}$ values, and also the low $m_0$ and $m_{1/2}$
(bulk) region of parameter space has become more accessible to direct DM 
searches and even indirect DM searches in the $\bar{p}$ channel. The CERN 
LHC can still cover the entire stau co-annihilation region, and the
low $m_{1/2}$ portion of the HB/FP region. The LC1000 can cover much of the 
stau co-annihilation region and much of the high $m_{1/2}$ HB/FP region.
However, a search of the $m_{1/2}<1400$ GeV portion of the 
HB/FP region can
be made by the IceCube neutrino telescope, 
and later, Stage 3 direct DM search experiments can cover the entire region. 

\FIGURE[htb]{
\epsfig{file=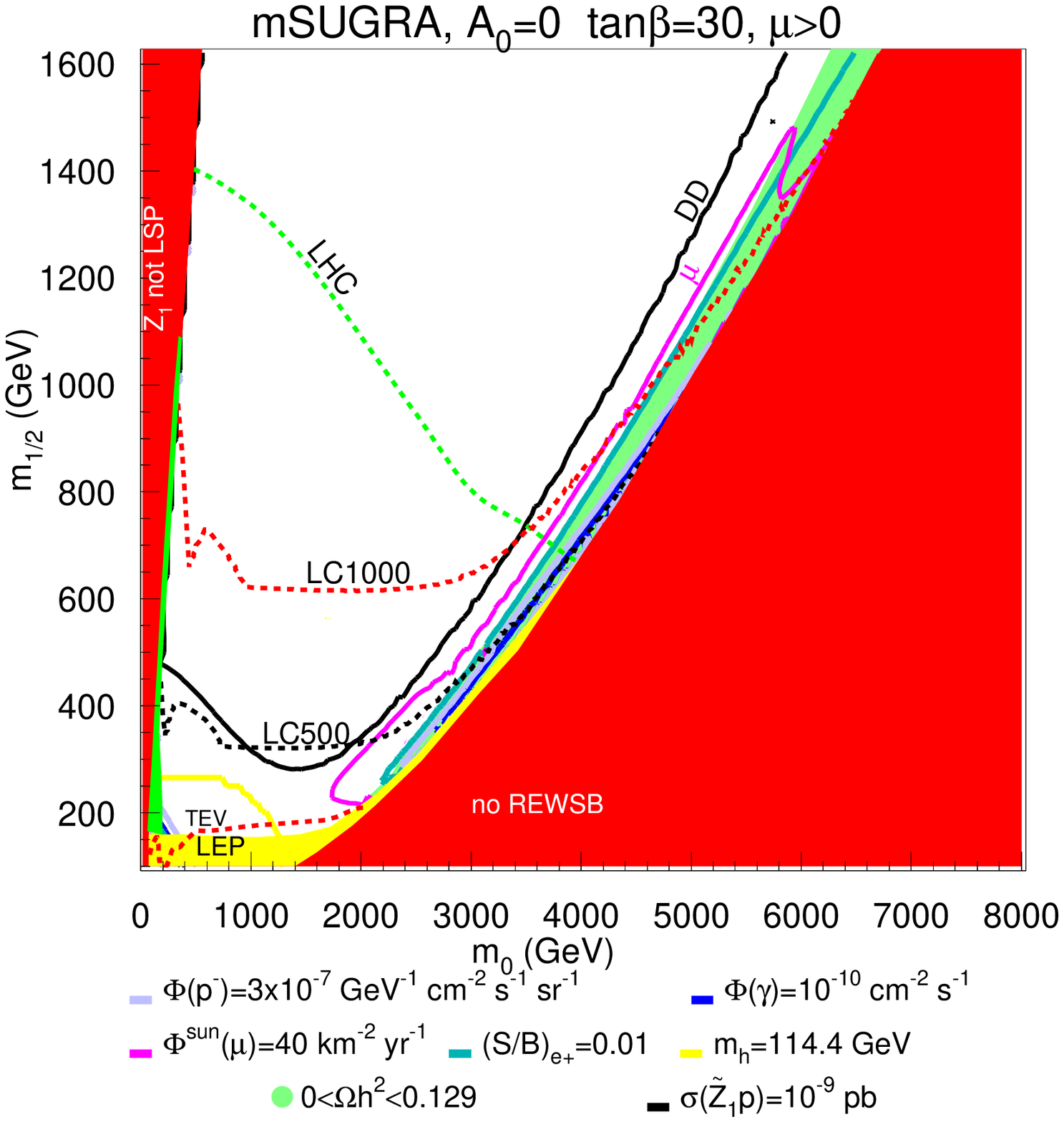,width=16cm} 
\caption{\label{fig30p}
A plot of the reach of direct, indirect and collider searches for 
neutralino dark matter in the $m_0\ vs.\ m_{1/2}$ plane, for 
$A_0=0$, $\tan\beta =30$ and $\mu >0$.}}

Fig. \ref{fig52p} shows the $m_0\ vs.\ m_{1/2}$ plane for an 
even higher $\tan\beta =52$ value. 
In this case, the stau co-annihilation corridor has increased to well 
beyond the LHC reach, and in fact this region of mSUGRA parameter space
appears to be one which is consistent with WMAP relic density
bounds, but beyond reach of {\it any} of the planned collider, 
direct or indirect
search experiments for neutralino dark matter. At this high of a 
$\tan\beta$ value, the value of pseudoscalar Higgs mass 
$m_A$ has dropped\cite{ltanb}
to such a level that there exists a large amplitude for off mass shell
$\tz_1\tz_1\to A^*\to b\bar{b}$ annihilation in the early universe\cite{bb2}.
This additional amplitude helps to amplify the low $m_0$ DM allowed
region, where now a combination of slepton co-annihilation, 
off shell $A,\ H$ resonance annihilation, and $t$-channel 
neutralino annihilation via relatively light sfermions all serve to 
expand the DM allowed region. However, the HB/FP region is relatively 
insensitive to changes in $\tan\beta$, and is still well covered by
$\nu$-telescopes and direct DM searches. The reach of direct DM search 
experiments has vastly increased in the low $m_0$ and $m_{1/2}$ regions,
where the reach of Stage 3 detectors is comparable to the LC1000.
The increase in direct DM detection rates is due to enhanced
scattering via $t$-channel Higgs exchange graphs; these are 
proportional to the square of $b$ and $\tau$ Yukawa couplings, which are
large at large $\tan\beta$\cite{bb2}.
In addition, at low $m_0$ and $m_{1/2}$, the DM allowed region may give rise 
to detectable rates for $\gamma$, $\bar{p}$ and $e^+$ detection.

\FIGURE[htb]{
\epsfig{file=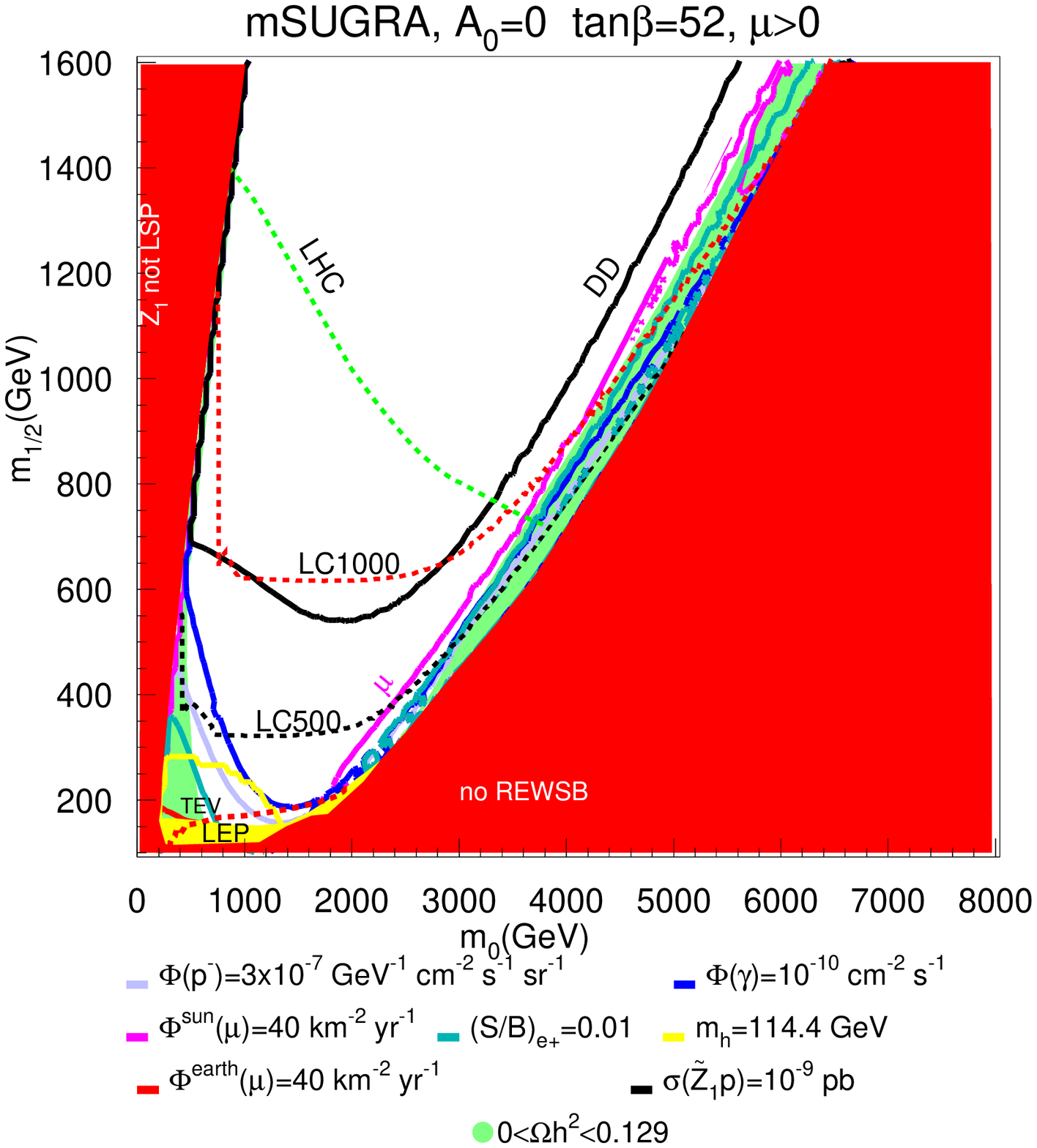,width=16cm} 
\caption{\label{fig52p}
A plot of the reach of direct, indirect and collider searches for 
neutralino dark matter in the $m_0\ vs.\ m_{1/2}$ plane, for 
$A_0=0$, $\tan\beta =52$ and $\mu >0$.}}

Fig. \ref{fig55p} shows the $m_0\ vs.\ m_{1/2}$ plane for
$\tan\beta =55$. In this case, the $A$ annihilation funnel has moved into
the central part of parameter space, opening up a large new region that 
gives a neutralino relic density in accord with WMAP results\cite{Afunnel}.
The $A$ annihilation funnel extends beyond the 100 fb$^{-1}$ reach of the
LHC, and makes a case for LHC running with much higher integrated 
luminosities, if no physics beyond the SM is found. In addition, the
searches for $\gamma$s, $e^+$s and $\bar{p}$s are all enhanced in ths region, 
since now halo neutralinos can also annihilate through the broad
$A$ and $H$ resonances\cite{bo}. Searches in these channels, however, 
cover only the low $m_{1/2}$ portion of the $A$ annihilation funnel.
Annihilation through the $A$ funnel also serves to expand somewhat
the breadth of the HB/FP region. Even so, most of the HB/FP region
can still be covered by the IceCube $\nu$ telescope, while
Stage 3 DM detectors can cover the entire region. 
Much of it can also be covered by LHC, LCs
and the search for halo annihilations into $\gamma$s, $e^+$s and $\bar{p}$s.
As $\tan\beta$ increases much beyond 55, the parameter space starts to
collapse due to inappropriate breakdown of EW symmetry.

\FIGURE[htb]{
\epsfig{file=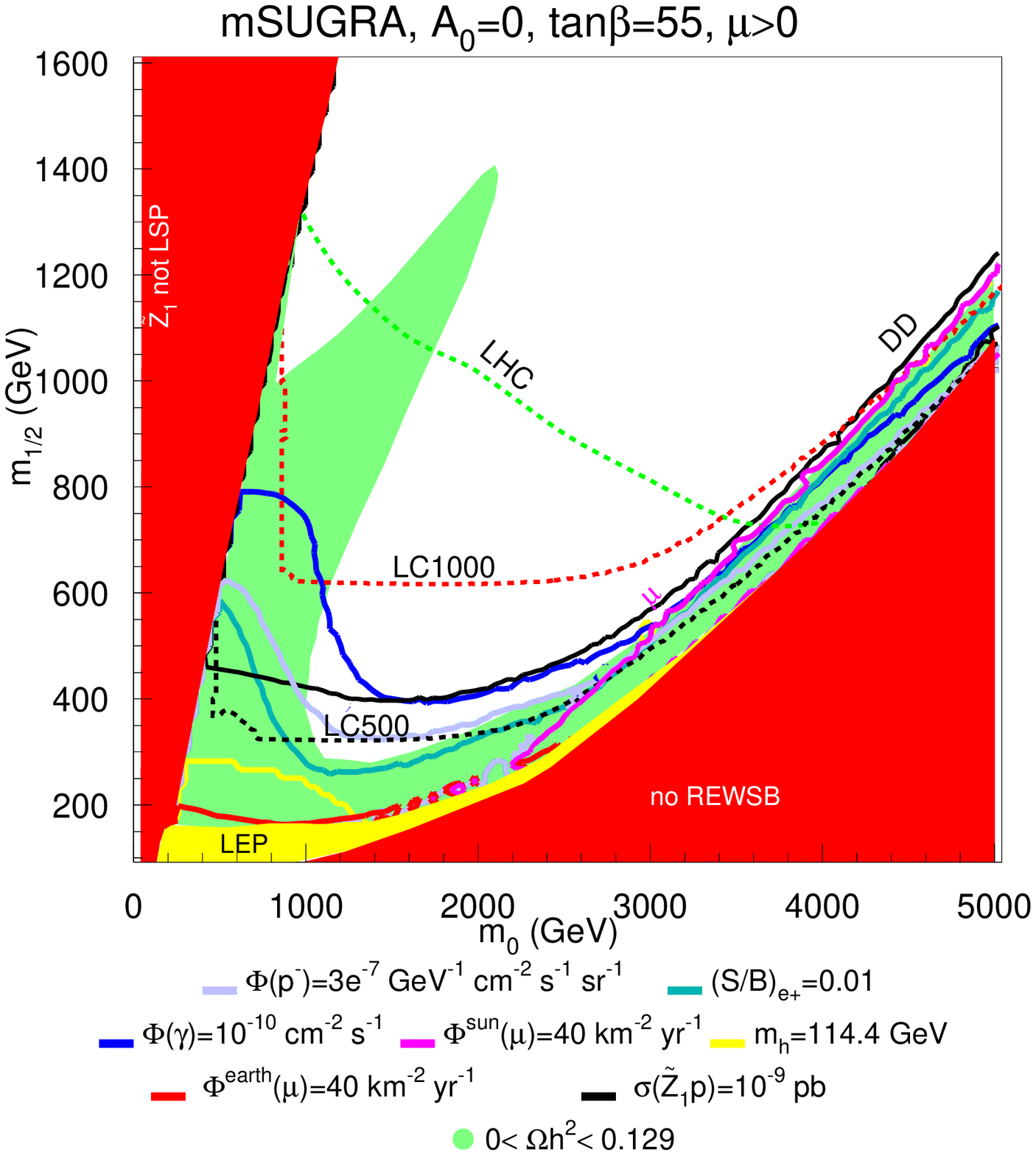,width=16cm} 
\caption{\label{fig55p}
A plot of the reach of direct, indirect and collider searches for 
neutralino dark matter in the $m_0\ vs.\ m_{1/2}$ plane, for 
$A_0=0$, $\tan\beta =55$ and $\mu >0$.}}

In Fig. \ref{fig45n}, we show the $m_0\ vs.\ m_{1/2}$ plane
for $\mu <0$, $A_0=0$ and $\tan\beta =45$. For negative $\mu$ values,
the $A$-annihilation funnel enters the plane at lower $\tan\beta$ values;
it is hence narrower than in Fig. \ref{fig55p}, since the $b$ and $\tau$
Yukawa couplings are smaller, and the $A$ and $H$ widths not so wide.
In this case, the CERN LHC covers almost all the stau co-annihilation region
and the $A$ funnel, and would certainly cover all of these with a higher
integrated luminosity. As before, the HB/FP region is only partially covered
by LHC and also by a LC, but again it is covered by the IceCube 
$\nu$ telescope up to $m_{1/2}\sim 1400$ GeV, and covered completely by 
Stage 3 direct DM detectors. In addition, the
enhanced rates for $\gamma$s, $e^+$s and $\bar{p}$s are 
displayed in the $A$ annihilation funnel.

\FIGURE[htb]{
\epsfig{file=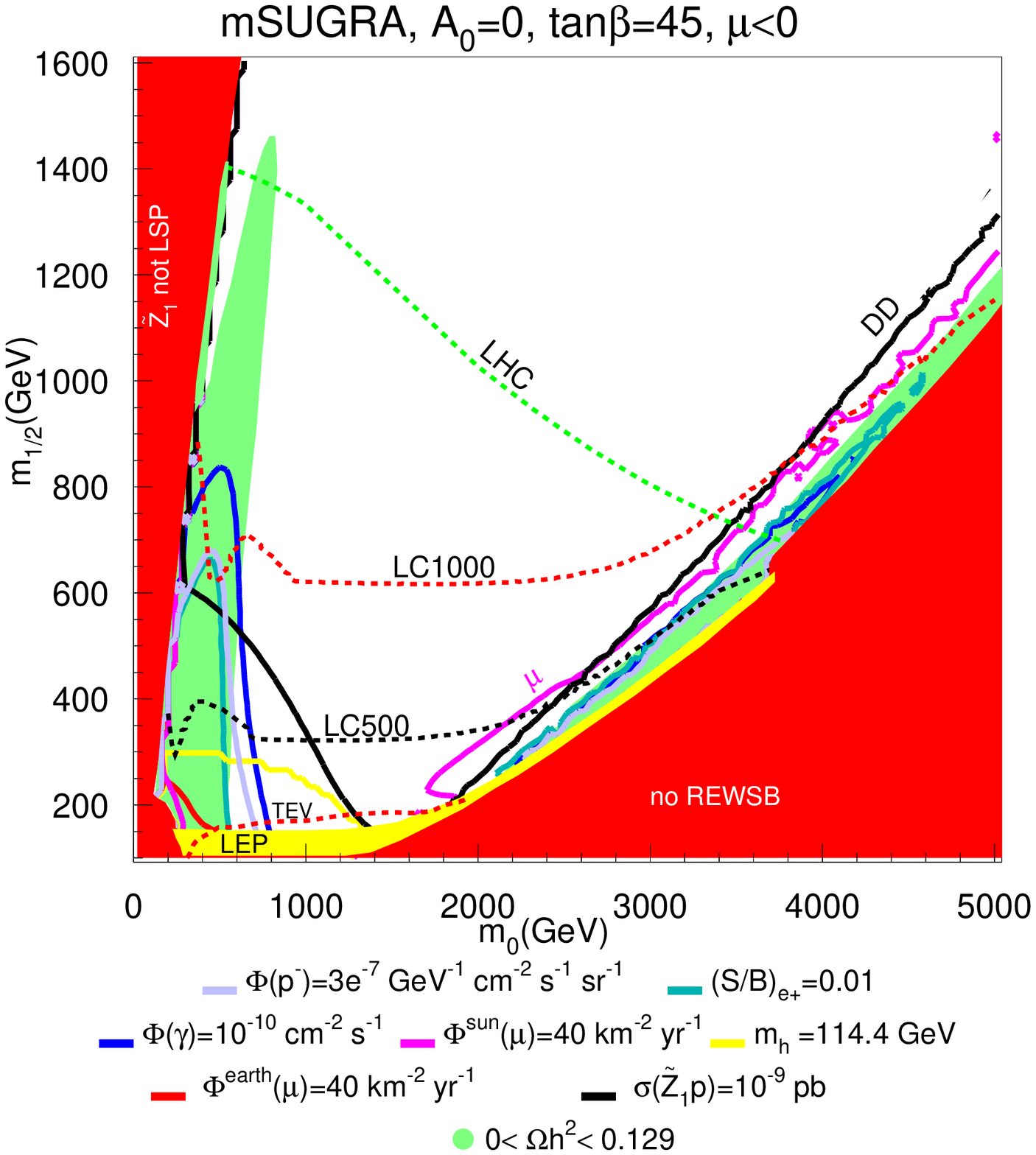,width=16cm} 
\caption{\label{fig45n}
A plot of the reach of direct, indirect and collider searches for 
neutralino dark matter in the $m_0\ vs.\ m_{1/2}$ plane, for 
$A_0=0$, $\tan\beta =45$ and $\mu <0$.}}

We show in Fig. \ref{fig50n} the same $m_0\ vs.\ m_{1/2}$ plane for
$\mu <0$, but this time for $\tan\beta =50$. In this case, the 
$A$ annihilation funnel is more centrally located, and shows
observable rates for $\gamma$s, $e^+$s and $\bar{p}$s for the
lower portion of the funnel. The red bulge moving into the figure at 
low $m_0$ and low $m_{1/2}$ denotes the region where $m_A^2<0$, 
and begins the collapse of parameter space at high $\tan\beta$.
The HB/FP region is again almost completely covered by IceCube, 
although the low $m_{1/2}$ 
portion of this region is no longer accessible to direct DM 
searches, owing to interferences in the neutralino-proton scattering rates.

\FIGURE[htb]{
\epsfig{file=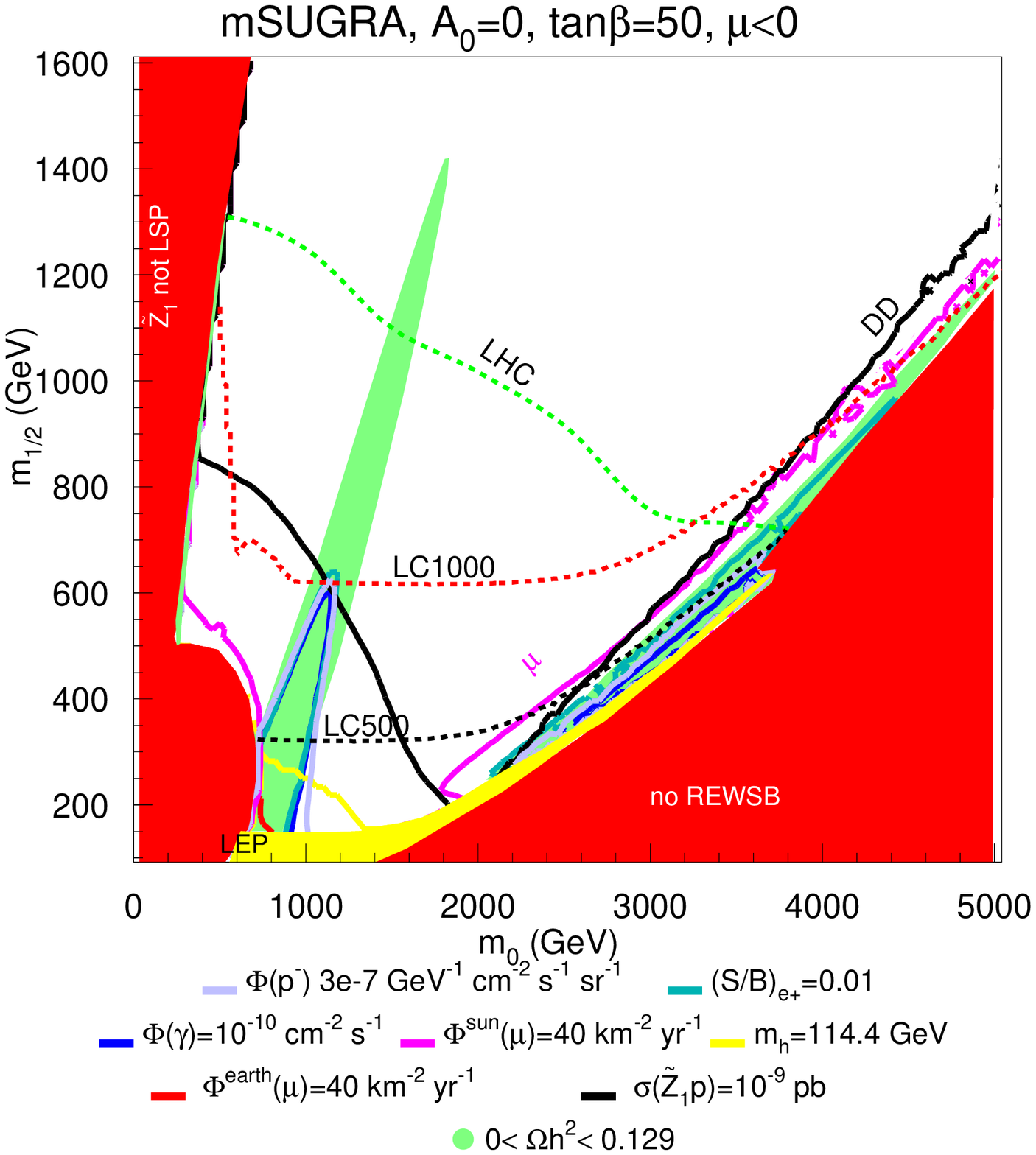,width=16cm} 
\caption{\label{fig50n}
A plot of the reach of direct, indirect and collider searches for 
neutralino dark matter in the $m_0\ vs.\ m_{1/2}$ plane, for 
$A_0=0$, $\tan\beta =50$ and $\mu <0$.}}
%


\section{Conclusions}
\label{conclude}

In previous reports, the reach of the Fermilab Tevatron, the CERN LHC 
and $\sqrt{s}=0.5$ and 1 TeV $e^+e^-$ linear colliders has been 
computed in the mSUGRA model. In addition, the reach of direct DM detection
experiments has also been worked out, and found to be in many respects 
complementary to collider searches\cite{bbbo}. 
In this paper, we augment these
previous works by presenting as well the reach of various indirect
search experiments for neutralino dark matter. These include searches for
neutralino annihilation in the core of the sun (or earth), leading to
detection of $\nu_\mu\to\mu$ conversions in neutrino telescopes
such as Antares and IceCube. Also, we show reach contours for
indirect searches for neutralino annihilation in the galactic core
via $\gamma$ detection, and for neutralino annihilation in the galactic
halo via $e^+$ and $\bar{p}$ detection. Our results have focussed
mainly on the WMAP allowed regions of the paradigm mSUGRA model. 

We have several main conclusions:
\begin{enumerate}
\item In the stau co-annihilation region, indirect searches
for neutralino dark matter have only a feeble reach. The best reach comes 
if $\tan\beta$ is large and there is some overlap with the 
$A$ annihilation funnel. However, the CERN LHC can probe {\it all}
the stau coannihilation region for $\tan\beta \alt 45$. 
Much of it can also be explored by linear $e^+e^-$ colliders.
In fact, the large $m_{1/2}$ portion of this region at large $\tan\beta$
seems to be a region of mSUGRA parameter space which is consistent with
WMAP limits on $\Omega_{\tz_1}h^2$, but not accessible to any planned 
experiments.
\item Most of the $A$ annihilation funnel can be explored by the CERN LHC,
although again the large $m_{1/2}$ portion of it might not be accessible
to any search experiments. The lower portion of the $A$ funnel is
also accessible to $\gamma$, $e^+$ and $\bar{p}$ searches for
neutralino annihilation in the galactic core or halo. The indirect detection 
reach in this region is comparable to that of a $\sqrt{s}=1$ TeV linear 
$e^+e^-$ collider. Detection of $\nu_\mu \to\mu$ in neutrino telescopes
is {\it unlikely} to occur in this region.
\item In the HB/FP region, the CERN LHC can cover $m_{1/2}$ values
up to $\sim 700$ GeV, corresponding to a value of $m_{\tg}\sim 1.8$ TeV.
Linear $e^+e^-$ collider can do better, since they can detect 
chargino pair production, even if the energy release from chargino
3-body decay is very low. The LC reach is limited by their CM energy,
and a 1 TeV linear collider will not be quite sufficient to explore
the entire high $m_{1/2}$ portion of the HB/FP region. However, 
in this region, rates for detection of neutrinos arising from 
neutralino annihilation in the core of the sun are large, and it seems
likely that IceCube can explore or rule out the HB/FP region
for $m_{1/2}<1400$ GeV.
In addition, Stage 3 direct DM detection experiments 
sensitive to neutralino proton spin-independent scattering cross sections
of $10^{-9}$ pb should be able to access the entire HB/FP region, unless
$\mu <0$ and $\tan\beta$ is large, in which case a small hole arises in the
low $m_{1/2}$ portion (which will be explored by LHC and LC anyway).
\end{enumerate}

Ultimately, the search for neutralino dark matter can proceed
via direct DM searches, indirect DM searches and collider searches.
By combining results, all the different search experiments
can cover almost all the WMAP allowed mSUGRA model parameter space, 
save for a few regions which occur in the high $m_{1/2}$ portion of the 
stau coannihilation corridor or the $A$ annihilation funnel.

{\it Note added:} A similar study of indirect and direct signals for 
neutralino dark matter in the mSUGRA model appeared shortly before the release 
of this work by Edsj\"o, Schelke and Ullio\cite{esu}. Where the two papers
overlap, we seem to be in agreement. 

\section*{Acknowledgments}
 
We thank X. Tata for conversations.
This research was supported in part by the U.S. Department of Energy
under contract number DE-FG02-97ER41022.
	
%

\end{document}